\documentclass[preprint]{iucr}              
\usepackage{graphicx}
     \paperprodcode{a000000}      
     \paperref{xx9999}            
     \papertype{FA}               
                                       \paperlang{english}          
     \journalcode{A}              
                                  
     \journalyr{2001}
     \journaliss{1}
     \journalvol{57}
     \journalfirstpage{000}
     \journallastpage{000}
     \journalreceived{0 XXXXXXX 0000}
     \journalaccepted{0 XXXXXXX 0000}
     \journalonline{0 XXXXXXX 0000}

\newcommand{\etal}{{\it et al.\/}\ }

\begin{document}                  



\title{Reconstruction of an object from its symmetry-averaged diffraction pattern}

\cauthor{Veit}{Elser}{ve10@cornell.edu}

\aff{Department of Physics, Cornell University, Ithaca, NY 14853-2501 \country{USA}}

\shortauthor{Elser}
     
\maketitle

\begin{synopsis}
An object can be reconstructed from its diffraction pattern even if the latter is incoherently averaged with respect to a discrete group of symmetries.
\end{synopsis}

\begin{abstract}
By suitably generalizing the Fourier constraint projection in the difference map phasing algorithm, an object can be reconstructed from its diffraction pattern even when the latter has been incoherently averaged over a discrete group of symmetries. This resolves an ambiguity in the recent proposal for aligning molecules by means of their anisotropic dielectric interaction with an intense light field. The algorithm is demonstrated with simulated data in two and three dimensions.
\end{abstract}

\section{Introduction}

If one is primarily interested in molecules, then crystals are merely a convenient alignment mechanism for improving the signal in a diffraction experiment. The near perfect alignment of molecules within a crystal, however, comes at a cost. In a crystal, alignment is achieved by introducing translational periodicity as well, and this restricts the sampling of the diffraction pattern to the crystal's reciprocal lattice. A more complete sampling of the diffraction pattern, also called \textit{oversampling}, is desirable when retrieving the phases in the pattern. This has led to the suggestion of alternate alignment mechanisms (Spence \& Doak, 2004), where not only would the molecules lack translational periodicity, but by arranging the molecular separations to be outside the range of coherence of the illuminating radiation, all inter-molecular interference would be eliminated.
If such an ``incoherent alignment" mechanism were realized, the scattered radiation would form the diffraction pattern of an individual molecule and be signal enhanced in proportion to the number of molecules in the illuminating beam.

One of the leading candidates for an incoherent alignment mechanism is the interaction of an elliptically polarized light field with the anisotropic molecular polarizability (Larsen \etal, 2000). As already noted (Spence \etal, 2005), this scheme suffers from the flaw that the energetics of the molecular orientation is degenerate with respect to a discrete symmetry group of order four generated by $\pi$-rotations about the principal axes of the polarizability tensor. The scattered radiation from many molecules will therefore be an average of four, generally distinct diffraction patterns, all related by symmetry. An incoherently averaged pattern will have no simple (linear) relationship to the scattering density: in particular, it does not correspond to the symmetry average of the scattering density. Incoherence proves to be a blessing here as well, however. As we show below, the molecule's scattering density can be reconstructed from a symmetry-averaged diffraction pattern by a relatively straightforward modification of a projection-based phase retrieval algorithm.

Not surprisingly, reconstruction from a symmetry-averaged diffraction pattern places greater demands on the degree of oversampling of the data. This should not pose a problem for the proposed alignment mechanism when applied to molecules, however, since the oversampling is then not limited by the transverse coherence of the illuminating beam, but rather, the time needed to acquire the signal. 

\section{Difference map algorithm}

We use the difference map algorithm (Elser, 2003a \& 2003b), a general purpose iterative method for finding a point in a Euclidean space that lies in the intersection of two constraint sets. The key components of the algorithm are the two projections, that given an arbitrary input, return as outputs the nearest points on the corresponding constraint sets. For the reconstructions considered here, the first projection implements the support and positivity constraint: 
\begin{equation}
\Pi_{S+}: \rho_{\mathbf{r}}\mapsto {\rho_{\mathbf{r}}}^\prime=
\left\{
\begin{array}{rl}
0 & \mbox{if $\mathbf{r}\notin S$ or $\rho_{\mathbf{r}}<0$,}\\
\rho_{\mathbf{r}} & \mbox{otherwise.}
\end{array}\right.
\end{equation}
Here $\rho_{\mathbf{r}}$ is the density at the direct space point $\mathbf{r}$ and $S$ is the known or estimated support of the scattering density. 

The second projection implements the constraint provided by the diffraction data and is most easily expressed in terms of the density in Fourier space, $\rho_{\mathbf{q}}$. Since we have a discrete group $G$ that averages the diffraction pattern, we consider the action of this projection on an orbit of densities $\{\rho_{g\mathbf{q}}\}_{g\in G}$ :
\begin{equation}
\Pi_{F}: \rho_{g\mathbf{q}}\mapsto {\rho_{g\mathbf{q}}}^\prime\; .
\end{equation}
The constraint is given by
\begin{equation}\label{constraint}
\sum_{g\in G}|{\rho_{g\mathbf{q}}}^\prime|^2=I_{G\mathbf{q}}\; ,
\end{equation}
where $I_{G\mathbf{q}}$ is the diffraction intensity associated with the orbit $G\mathbf{q}$. The output of $\Pi_{F}$ satisfies (\ref{constraint}) while also minimizing the distance
\begin{equation}\label{dist}
\|\rho-\rho^\prime\|^2=\sum_{G\mathbf{q}}\;\sum_{g\in G}|{\rho_{g\mathbf{q}}}-{\rho_{g\mathbf{q}}}^\prime|^2\; .
\end{equation}

The derivation of a formula for the projection $\Pi_{F}$ proceeds in two steps. First, consider the nonnegative amplitudes
\begin{eqnarray}
f_{g\mathbf{q}}&=&|\rho_{g\mathbf{q}}|\\
{f_{g\mathbf{q}}}^\prime&=&|{\rho_{g\mathbf{q}}}^\prime|\label{fprime}\; .
\end{eqnarray}
For any value of ${f_{g\mathbf{q}}}^\prime$ (to be determined later), the complex Fourier amplitude ${\rho_{g\mathbf{q}}}^\prime$ that minimizes (\ref{dist}) subject to (\ref{fprime}) is given by the projection to the circle:
\begin{equation}\label{tocircle}
{\rho_{g\mathbf{q}}}^\prime=\left(\frac{{f_{g\mathbf{q}}}^\prime}{f_{g\mathbf{q}}}\right)\rho_{g\mathbf{q}}\; .
\end{equation}
The distance (\ref{dist}) can now be expressed in terms of the amplitudes:
\begin{equation}\label{dist2}
\|\rho-\rho^\prime\|^2=\sum_{G\mathbf{q}}\;\sum_{g\in G}(f_{g\mathbf{q}}-{f_{g\mathbf{q}}}^\prime)^2\; .
\end{equation}
In the second step of the derivation we minimize (\ref{dist2}) with respect to ${f_{g\mathbf{q}}}^\prime$ subject to the constraint (\ref{constraint}) written as
\begin{equation}
\sum_{g\in G}({f_{g\mathbf{q}}}^\prime)^2=I_{G\mathbf{q}}\; .
\end{equation}
This is solved by projecting to the sphere, in analogy to the circle projection in the complex plane, only now the dimension of the sphere is one less the cardinality of the orbit $G\mathbf{q}$ (a divisor of the order of $G$):
\begin{equation}
{f_{g{\mathbf{q}}}}^\prime=\frac{\sqrt{I_{{G\mathbf{q}}}}}{\left(\sum_{g\in G}(f_{g{\mathbf{q}}})^2\right)^{1/2}}\;f_{g{\mathbf{q}}}\; .
\end{equation}
Combining this with (\ref{tocircle}) we obtain the formula
\begin{equation}
\Pi_{F}(\rho_{g\mathbf{q}})=\frac{\sqrt{I_{{G\mathbf{q}}}}}{\left(\sum_{g\in G}|\rho_{g{\mathbf{q}}}|^2\right)^{1/2}}\; \rho_{g{\mathbf{q}}}\;.
\end{equation}

Below we use the difference map with $\beta=1$, for which it reduces to a generalization of Fienup's input-output map (Fienup, 1982):
\begin{equation}
D:\rho\mapsto\rho^\prime=\rho+\Delta(\rho)
\end{equation}
\begin{equation}\label{delta}
\Delta(\rho)=\Pi_{S+}\left(2\,\Pi_F(\rho)-\rho\right)-\Pi_F(\rho)\; .
\end{equation}
When the diffraction pattern is not symmetry-averaged, the limiting oversampling ratio for real-valued objects is $\sigma>2$ (Miao \etal, 1998). With symmetry-averaging, the number of Fourier constraints is reduced by the order of the group and the limit becomes $\sigma>2 |G|$. For the alignment scheme considered above, $|G|=4$. 

\section{Numerical experiments}

The diffraction pattern of a real-valued object in two dimensions has inversion symmetry and results in a two-fold ambiguity in the reconstructed object. When the diffraction pattern is averaged with respect to a mirror, the ambiguity becomes four-fold and is generated by two orthogonal mirrors. We demonstrate this for the object ``R", one of the least symmetrical letters in the alphabet. Figure 1 shows the object in a $27\times 27$ pixel support, the diffraction pattern when expanded into a $128\times 128$ field of view, and the diffraction pattern averaged with respect to a vertical (or horizontal) mirror. The algorithm used a slightly enlarged, $30\times 30$, support constraint to recover the object. Figure 2 shows the time series of the 
difference map error metric $\|\Delta\|$ given by (\ref{delta}). When $\|\Delta\|$ vanishes, the density
\begin{equation}
\rho_\mathrm{sol}=\Pi_{S+}\left(2\,\Pi_F(\rho)-\rho\right)
\end{equation}
satisfies the support/positivity constraint but also the symmetry-averaged Fourier constraint, since
\begin{equation}
\Pi_{S+}\left(2\,\Pi_F(\rho)-\rho\right)=\Pi_F(\rho)\; .
\end{equation}
Also shown in Figure 2 is the reconstructed object; in this case the mirror of the original.

Our three dimensional reconstruction experiment reproduced the symmetry ambiguity of the proposed molecular alignment mechanism (Spence \etal, 2005). A real-valued density was generated by applying a rectifier (positivity projection) to a random, low-pass filtered density on a $12\times 12\times 12$ support. The diffraction pattern was computed for a $32\times 32\times 32$ field of view and averaged with respect to the four element group generated by $\pi$-rotations about axes along the sampling grid. When combined with the Friedel symmetry of the non-averaged diffraction pattern, the input to the reconstruction algorithm thus has an eight element symmetry group generated by the three coordinate reflections.

Figure 3 shows the times series of the difference map error metric for ten successive reconstructions from different random starts. The corresponding densities are shown row-wise in Figure 4. Of the eight possible symmetry related reconstructions, the ten trials produced six.

\ack{Acknowledgement}

This work was supported by Department of Energy grant DE-FG02-05ER46198.

\newpage

\begin{figure}
\caption{\textit{Left}: object of study in a 2D reconstruction experiment. \textit{Center}: diffraction pattern, \textit{right}: diffraction pattern averaged with its mirror.}
\centerline{\scalebox{1.5}{\includegraphics{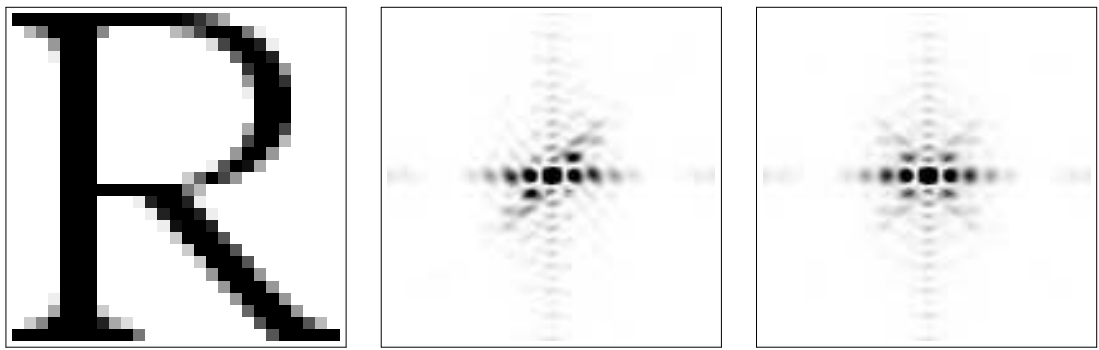}}}
\end{figure}

\begin{figure}
\caption{\textit{Left}: Time series of the difference map error metric $\|\Delta\|$ vs. iteration number; \textit{right}: reconstructed object.}
\centerline{\scalebox{1.5}{\includegraphics{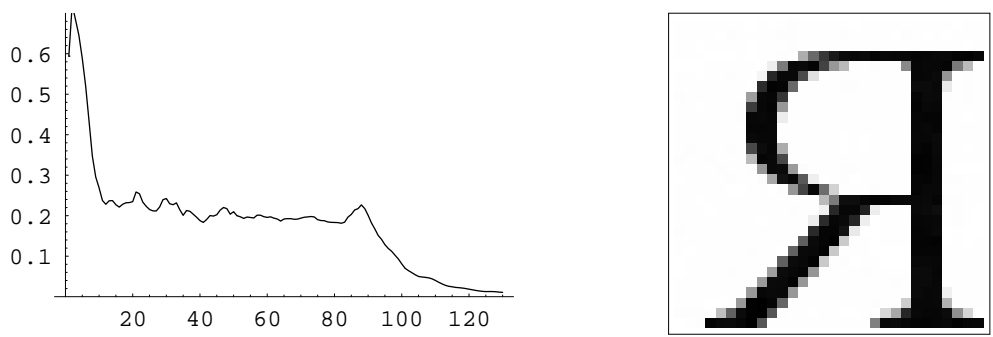}}}
\end{figure}

\newpage

\begin{figure}
\caption{Error metric time series for the ten 3D reconstructions shown in Figure 4. All solutions required fewer than 300 iterations.}
\centerline{\scalebox{1.5}{\includegraphics{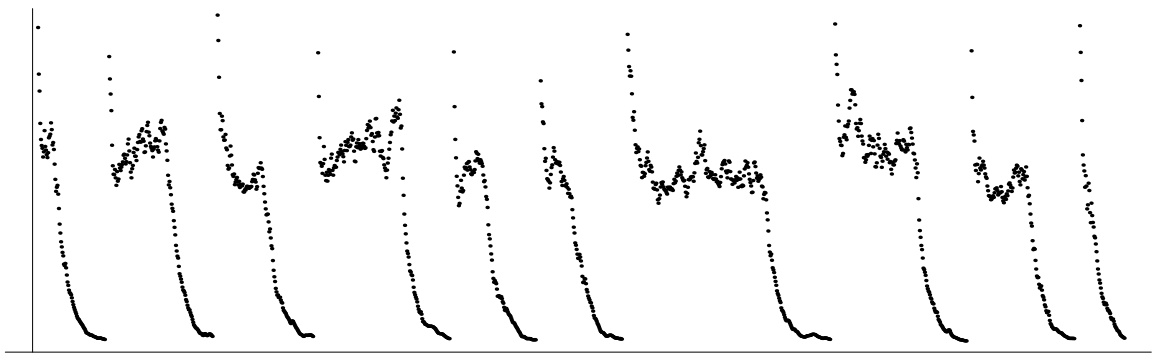}}}
\end{figure}

\begin{figure}
\caption{Each horizontal row of twelve, $12\times 12$ pixel arrays, represents one 3D reconstruction from symmetry averaged data. Of the eight symmetry related reconstructions possible, the ten experiments shown realize six of these (rows resorted to ease comparison).}
\centerline{\scalebox{1.5}{\includegraphics{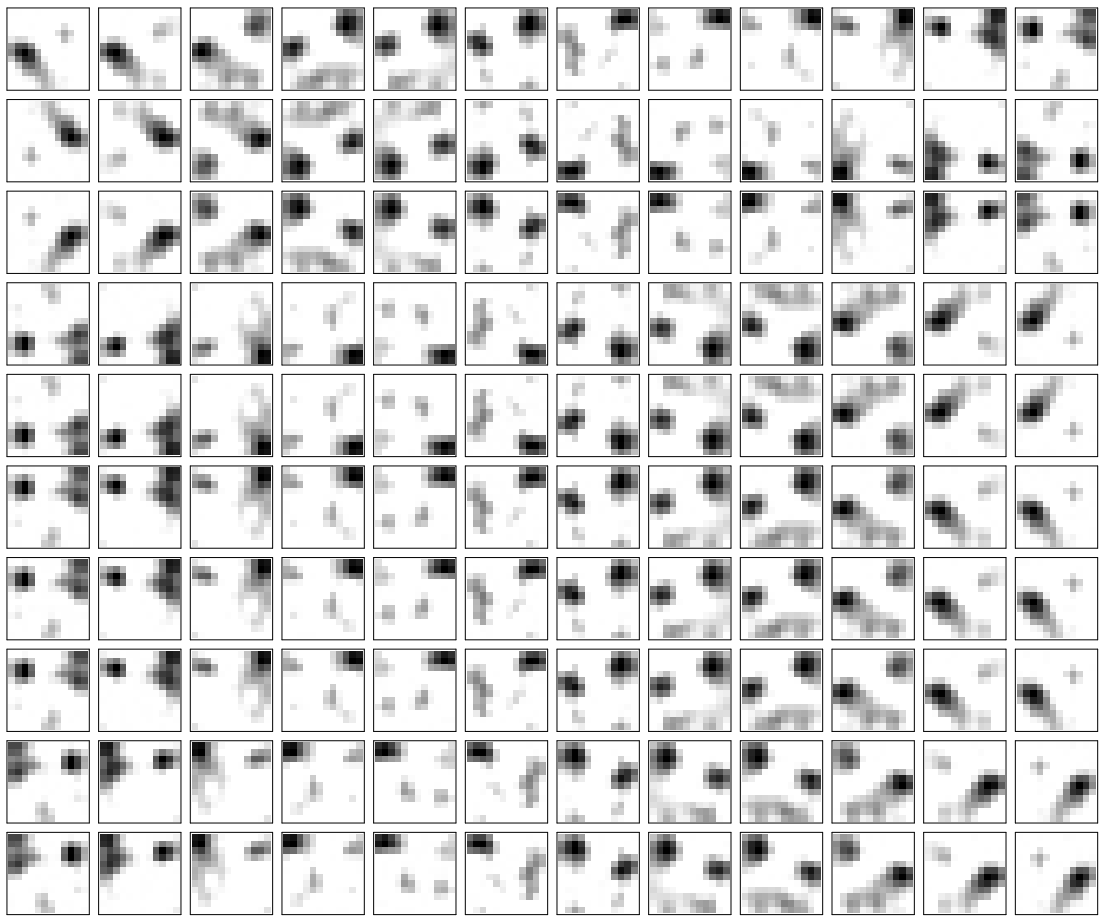}}}
\end{figure}

\end{document}